\documentclass[pdftex,twocolumn,epjc3]{svjour3}   
\RequirePackage[T1]{fontenc}

\smartqed

\usepackage{slashed}
\usepackage{dcolumn}
\usepackage{amsmath}
\usepackage{ulem}

\RequirePackage{graphicx}
\RequirePackage{flushend}
\RequirePackage[numbers,sort&compress]{natbib}
\RequirePackage[colorlinks,citecolor=blue,urlcolor=blue,linkcolor=blue]{hyperref}

\journalname{Eur. Phys. J. A}

\begin{document}

\title{Accessing weak neutral-current coupling $g_{AA}^{eq}$ using positron and electron beams at Jefferson Lab}
\author{Xiaochao Zheng\thanksref{e1,addr1} \and
        Jens Erler\thanksref{e2,addr2}$^\clubsuit$ \and
        Qishan Liu\thanksref{e3,addr2}$^\spadesuit$ \and
        Hubert Spiesberger\thanksref{e4,addr2}$^\spadesuit$}
\thankstext{e1}{e-mail: xiaochao@jlab.org}
\thankstext{e2}{e-mail: erler@uni-mainz.de}
\thankstext{e3}{e-mail: qisliu@students.uni-mainz.de}
\thankstext{e4}{e-mail: spiesber@uni-mainz.de}

\institute{{Department of Physics, University of Virginia, 
382 McCormick Rd, Charlottesville, VA 22904, USA}\label{addr1} \and
{PRISMA$^+$ Cluster of Excellence, Institute for Nuclear Physics$^\clubsuit$ and Institute of 
Physics$^\spadesuit$, \\ 
\hspace*{2pt} Johannes Gutenberg-University, 55099 Mainz, Germany}\label{addr2}}

\date{Received: date / Accepted: date}

\maketitle

\begin{abstract}
Low-energy neutral-current couplings arising in the Standard Model of electroweak interactions can be
constrained in lepton scattering off hydrogen or a nuclear fixed target. 
Recent polarized electron scattering experiments at Jefferson Lab (JLab) have improved the precision in
the parity-violating types of effective couplings. 
On the other hand, the only known way to access the parity-conserving counterparts is to compare scattering 
cross sections between a lepton and an anti-lepton beam.  
We review the current knowledge of both types of couplings and how to constrain them.
We also present exploratory calculations for a possible measurement of $g_{AA}^{eq}$ using the planned SoLID spectrometer 
combined with a possible positron beam at JLab. 
\end{abstract}

\section*{Weak Neutral-Current Couplings from Charged Lepton Scattering}
The Lagrangian of the weak neutral-current interaction relevant to electron deep inelastic scattering (DIS) 
off quarks inside the nucleon is given by~\cite{PDG:2020},
\begin{eqnarray}
L_{NC}^{eq} &=& \frac{G_F}{\sqrt{2}} \sum_q \left[ 
g_{VV}^{eq} \bar e\gamma^\mu e \bar q\gamma_\mu q + 
g_{AV}^{eq} \bar e\gamma^\mu\gamma_5 e \bar q\gamma_\mu q \right. \nonumber \\ 
&+& \left. g_{VA}^{eq} \bar e\gamma^\mu e \bar q\gamma_\mu \gamma_5 q + 
g_{AA}^{eq} \bar e\gamma^\mu \gamma_5 e \bar q\gamma_\mu \gamma_5 q \right], \label{eq:L}
\end{eqnarray}
where $G_F$ is the Fermi constant.
The $g_{VV}^{eq}$ terms are shown here for completeness, even though they are not usually included in equations 
like~(\ref{eq:L}) as their effects are completely overwhelmed by those of QED, 
at the very least at lower energies.
At lowest order (tree level) in the Standard Model (SM), the four-fermion couplings are products of the lepton and quark 
couplings $g_{V,A}^f$ to the $Z_0$ boson,
\begin{eqnarray}
g_{AV}^{eu} &=& 2 g_A^e g_V^u = - \frac{1}{2} + \frac{4}{3}\sin^2\theta_W\ , \label{eq:c1u} \\
g_{AV}^{ed} &=& 2 g_A^e g_V^d = + \frac{1}{2} - \frac{2}{3}\sin^2\theta_W\ , \label{eq:c1d} \\
g_{VA}^{eu} &=& 2 g_V^e g_A^u = - \frac{1}{2} + 2\sin^2\theta_W = - g_{VA}^{ed} = - 2 g_V^e g_A^d \label{eq:c2ud} \\
g_{AA}^{eu} &=& - 2 g_A^e g_A^u = + \frac{1}{2} = - g_{AA}^{ed} = 2 g_A^e g_A^d\ , \label{eq:c3ud}
\end{eqnarray}
where $\theta_W$ is the weak mixing angle. 
The couplings $g^{eq}_{AV}$, $g^{eq}_{VA}$, and $g^{eq}_{AA}$~\cite{Erler:2013xha} agree to lowest order
with the widely used quantities $C_{1q}$, $C_{2q}$, and $C_{3q}$, respectively, 
but are defined independently of the processes in which they are measured. 
Their precise numerical values differ from those of $C_{1,2,3}$'s because they absorb some higher-order radiative corrections and can also be modified by possible physics beyond the SM.

The $g_{AV}^{eq}$ and $g_{VA}^{eq}$ terms in Eq.~(\ref{eq:L}) induce parity violation and a cross section asymmetry 
between left- and right-handed electrons scattering off unpolarized protons and nuclei~\cite{Erler:2014fqa}. 
Specifically, the couplings $g_{AV}^{eq}$ have been measured in  
elastic scattering, while the combination $2 g_{VA}^{eu} - g_{VA}^{ed}$ enters the polarization 
asymmetry in deep-inelastic scattering (DIS). The coupling $g_{AV}^{eq}$ was also determined in atomic parity violation experiments. 
The terms involving the $g_{AA}^{eq}$ do not violate parity, but 
can be accessed by comparing 
cross sections of lepton to anti-lepton DIS, as we will discuss here.

\begin{figure}[t]
\includegraphics[width=0.47\textwidth]{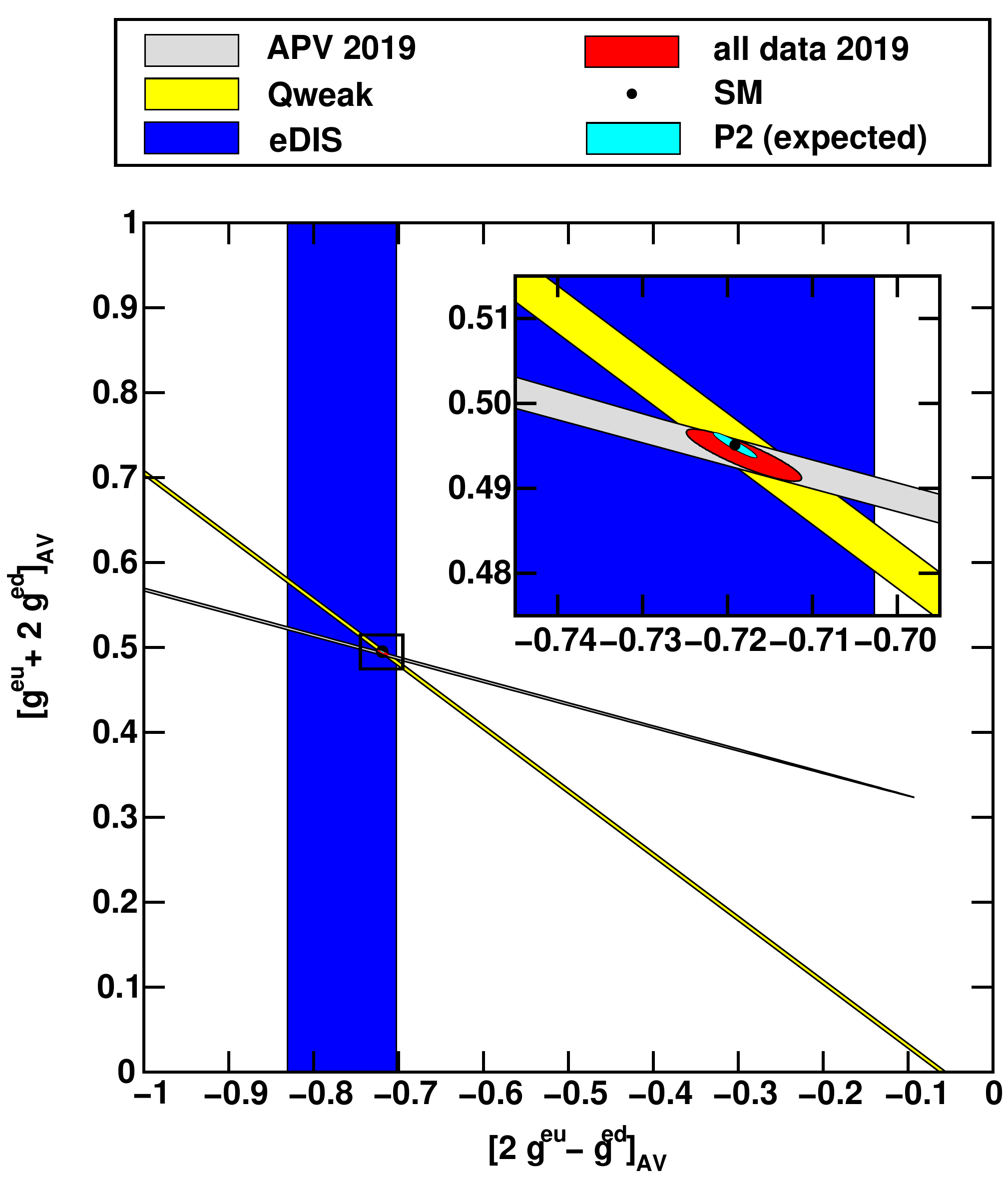}
\caption{Current experimental knowledge of the couplings $g_{AV}^{eq}$. 
The latest measurement is from the 6 GeV Qweak experiment~\cite{Androic:2018kni} at JLab. The Atomic Parity Violation ("APV 2019") results shown utilized the theory calculations of Ref.~\cite{Toh:2019iro}. The "eDIS" band is a combination of the SLAC E122~\cite{Prescott:1978tm,Prescott:1979dh} and the JLab PVDIS~\cite{Wang:2014bba,Wang:2014guo} experiments. 
Also indicated are the expected uncertainties from the planned P2 experiment~\cite{Becker:2018ggl} at Mainz,
centered at the SM value.
\label{fig:c1pm}}
\end{figure}

\begin{figure}[t]
\includegraphics[width=0.47\textwidth]{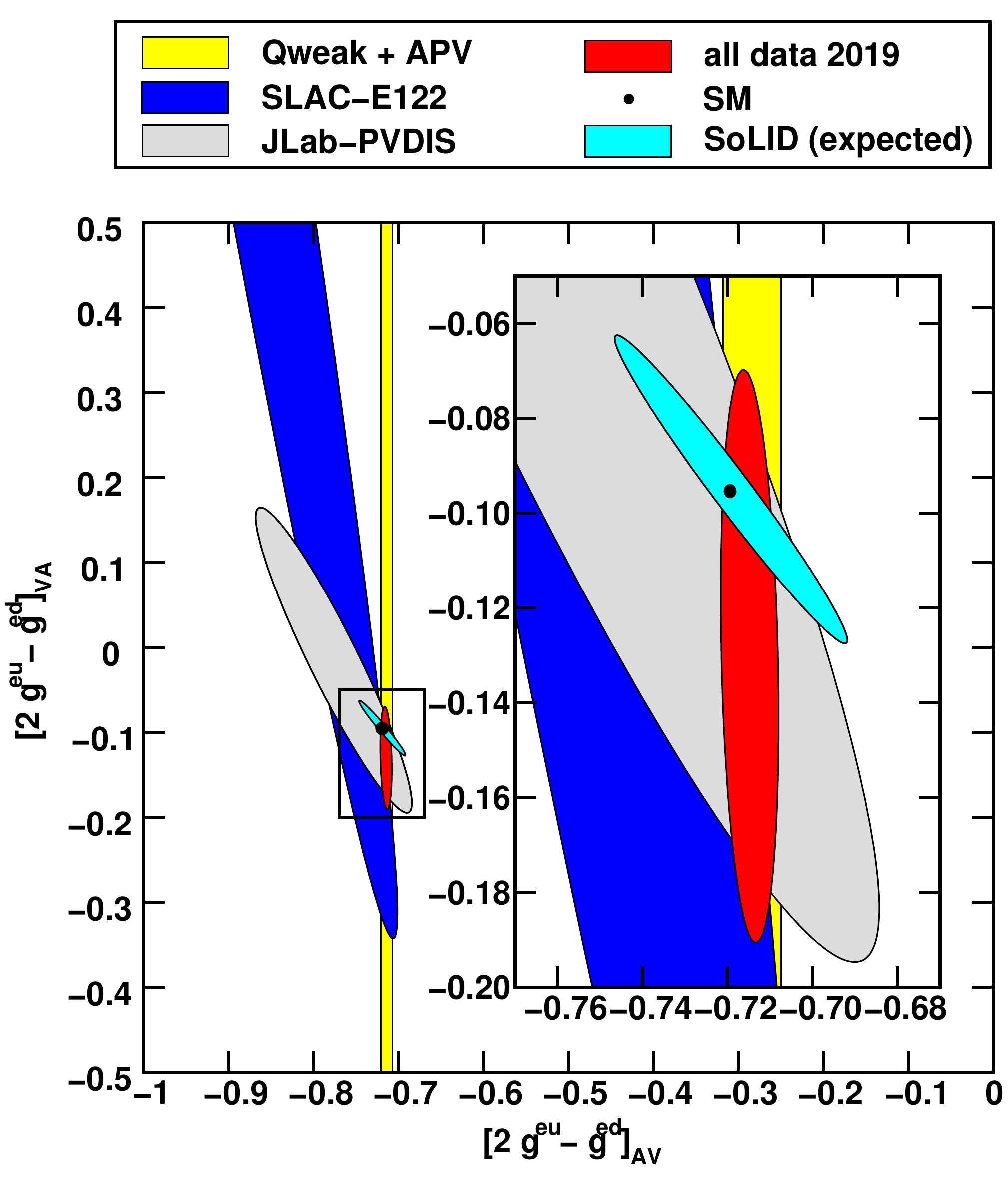}
\caption{Current experimental knowledge of the couplings $g_{VA}^{eq}$ in the combinations given 
by the electric charge ratio of up and down quarks.
The latest measurement is from the PVDIS experiment~\cite{Wang:2014bba,Wang:2014guo} at JLab. 
Also indicated are the expected uncertainties from the planned SoLID project~\cite{Chen:2014psa} at JLab, centered at the SM value. 
\label{fig:c2pm}}
\end{figure}

The parity-violating couplings can be determined in global fits.
Figure~\ref{fig:c1pm} presents $g_{AV}^{eu} + 2 g_{AV}^{ed}$ and similarly Fig.~\ref{fig:c2pm} shows 
$2 g_{VA}^{eu} - g_{VA}^{ed}$, in both cases as functions of the combination $2 g_{AV}^{eu} - g_{AV}^{ed}$.
The constraints in Fig.~\ref{fig:c1pm} are strongly dominated by measurements of atomic parity violation in
$^{133}$Cs~\cite{Wood:1997zq,Guena:2005uj} and parity-violating electron elastic scattering off 
protons~\cite{Androic:2018kni,Androic:2013rhu}.
The additional information entering Fig.~\ref{fig:c2pm} has been extracted from the parity-violating DIS 
experiments at SLAC~\cite{Prescott:1978tm,Prescott:1979dh}, as well as at JLab in the 6~GeV 
era~\cite{Wang:2014bba,Wang:2014guo}.

In contrast to the $g_{AV}^{eq}$ and $g_{VA}^{eq}$, direct measurement on the $g_{AA}^{eq}$ does not
yet exist.  
So far, there is only experimental information from CERN~\cite{Argento:1982tq} on their muonic counterparts,
obtained by means of comparing the DIS cross section of positively charged left-handed muons directed 
on a carbon target with that of negatively charged right-handed ones.  Neglecting radiative 
effects, their results can be written as 
\begin{eqnarray}
2g_{AA}^{\mu u}-g_{AA}^{\mu d} + 0.81\, (2g_{VA}^{\mu u}-g_{VA}^{\mu d}) &=& 1.45 \pm 0.41,\\
2g_{AA}^{\mu u}-g_{AA}^{\mu d} + 0.66\, (2g_{VA}^{\mu u}-g_{VA}^{\mu d}) &=& 1.70 \pm 0.79, 
\end{eqnarray} 
for the two beam energies $E_\mu = 200$~GeV and 120~GeV, and may be compared to 
the SM tree level predictions of $1.42$ and $1.44$, respectively. 
Note, that these results were previously summarized in Ref.~\cite{Erler:2004cx} but the calculations are 
updated here using $\alpha^{-1} = 129$ and $\alpha^{-1} = 130$ for the inverse of the electromagnetic coupling
at the two energies.

With the SM value for $2 g_{VA}^{\mu u} - g_{VA}^{\mu d} = - 0.0954$, 
which is in good agreement with the PVDIS measurement~\cite{Wang:2014bba,Wang:2014guo}, we find the constraint,
\begin{equation}
2g_{AA}^{\mu u} - g_{AA}^{\mu d} = 1.57 \pm 0.38, \label{eq:c3q_cern_value}
\end{equation}
where we assumed that the (smaller) systematic error of the 200~GeV data was common to both beam energies. 
Assuming lepton universality, one may compare the error in Eq.~(\ref{eq:c3q_cern_value}) with 
the uncertainties shown in Fig.~\ref{fig:c1pm} and Fig.~\ref{fig:c2pm}. 
But we stress that there is so far no direct measurement of the $g_{AA}^{eq}$ for electron-quark interactions. 

\section*{Neutral-Current Asymmetries in Lepton Scattering}
The electroweak neutral-current induces various kinds of non-vanishing asymmetries for lepton and anti-lepton
scattering, such as,
\begin{equation}
A^{e^+e^-}_{ij} \equiv \frac{\sigma_i^{e^+}-\sigma_j^{e^-}}{\sigma_i^{e^+}+\sigma_j^{e^-}}\ , \hspace{24pt}
A^{e^\pm}_{RL} \equiv \frac{\sigma_R^{e^\pm}-\sigma_L^{e^\pm}}{\sigma_R^{e^\pm}+\sigma_L^{e^\pm}}\ ,
\label{eq:Asym}
\end{equation}
with $i,j = R,L$, and analogous asymmetries for the heavier charged leptons.
These are related to the asymmetries appearing in Ref.~\cite{Spiesberger:1989rp} by,
\begin{align}
&A_\pm = - A^{e^\pm}_{RL}\ , \hspace{24pt} &C_i = A^{e^+e^-}_{ii}\ , \\
&B_+ = A^{e^+e^-}_{RL}\ , \hspace{24pt} &B_- = A^{e^+e^-}_{LR}\ .
\end{align}
Additionally, for unpolarized beams one can define,
\begin{eqnarray}
A^{e^+e^-}&\equiv&\frac{\sigma^{e^+}-\sigma^{e^-}}{\sigma^{e^+}+\sigma^{e^-}}\ ,
\label{eq:Apm_unpol}
\end{eqnarray}
which is related to $A_{RL}^{e^+e^-}$, see Ref.~\cite{Berman:1973pt}. 

$A^{e^-}_{RL}$ was first measured at SLAC~\cite{Prescott:1978tm,Prescott:1979dh} in DIS,
and then more precisely at JLab~\cite{Wang:2014bba,Wang:2014guo}, while $A^{\mu^+\mu^-}_{LR}$ is the 
aforementioned asymmetry as measured at CERN. 
Note that experiments targeting $A^{e^+e^-}$ have a great advantage for the positron beam being considered at JLab~\cite{Cardman:2018svy}, in that much higher luminosities are 
achievable when polarization is not required.
We provide explicit derivations of the asymmetries in Eqs.~(\ref{eq:Asym}) in the appendix. 

We now focus on asymmetries between positron and electron scattering, ignoring for simplicity nuclear and 
higher-order corrections. 
Considering the four quark flavors $u,d,c,s$ and assuming symmetric charm and strange seas, 
$c=\bar c$ and $s=\bar s$, we have for isoscalar targets such as the deuteron,
\begin{eqnarray}
A^{e^+e^-}_{RL,d} &=& \frac{3 G_F Q^2}{2\sqrt{2}\pi\alpha} Y(y) \frac{R_V}{5 + 4 R_C + R_S} \nonumber \\
&\times& \Big[ \vert\lambda\vert (2g_{VA}^{eu}-g_{VA}^{ed})-(2g_{AA}^{eu}-g_{AA}^{ed}) \Big]\ , \label{Aepem}
\end{eqnarray}
where $\vert\lambda\vert$ is the magnitude of the beam polarization.
$Q^2 \equiv -q^2$ with $q$ the 4-momentum transfer from the beam to the target, and $y$ is the fractional 
energy transfer.
In terms of parton distribution functions (PDFs) with their dependence on Bjorken $x$ and $Q^2$ implicit,
and abbreviating $q^+ \equiv q+ \bar q$ and $q_V \equiv q- \bar q$, one has,
\begin{align}
&Y(y)\equiv\frac{1-(1-y)^2}{1+(1-y)^2}\ , \hspace{24pt}
&R_V \equiv \frac{u_V+d_V}{u^+ + d^+}\ , \label{eq:Y} \\
&R_C \equiv \frac{2(c+\bar c)}{u^+ + d^+}\ , \hspace{24pt}
&R_S \equiv \frac{2(s+\bar s)}{u^+ + d^+}\ ,
\end{align}
and similarly, 
\begin{eqnarray}
A^{e^+e^-}_{RR,d} &=& \frac{3 G_F Q^2}{2\sqrt{2}\pi\alpha(5 + 4 R_C + R_S)}
\left\{ -\vert\lambda\vert\ [2 (1 + R_C) g_{AV}^{eu} \right. \nonumber \\
&-& \left. (1 + R_S) g_{AV}^{ed}] - Y(y) R_V (2 g_{AA}^{eu} - g_{AA}^{ed}) \right\}. \label{eq:ARR-d0}
\end{eqnarray}
The expressions~(\ref{Aepem})
and (\ref{eq:ARR-d0}) can also be applied to $A^{e^+e^-}_{LR,d}$ and $A^{e^+e^-}_{LL,d}$, 
respectively, provided that the sign $\vert\lambda\vert\to -\vert\lambda\vert$ is flipped.

For unpolarized beams, $\vert\lambda\vert=0$, Eq.~(\ref{Aepem}) and (\ref{eq:ARR-d0}) obviously simplify,
\begin{align} 
A^{e^+e^-}_d &= -\frac{3G_F Q^2}{2\sqrt{2}\pi\alpha} Y(y) 
\frac{R_V\left(2g_{AA}^{eu}-g_{AA}^{ed}\right)}{5 + 4 R_C + R_S} \nonumber \\
&= - 1.06 \times 10^{-4}\, Q^2\, \frac{Y(y) R_V (2g_{AA}^{eu}-g_{AA}^{ed})}{1+0.8\, R_C + 0.2\, R_S}\ ,
\label{eq:asym}
\end{align}
where $Q^2$ is in GeV$^2$, and where in the second line we assumed $\alpha^{-1} = 134$.
From Eq.~(\ref{eq:asym}) one can see that the first generation $g_{AA}^{eq}$ can be measured directly by comparing
$e^-$ and $e^+$ DIS cross sections, ideally with a high-intensity positron beam.
Furthermore, to isolate the $g_{AA}^{eq}$ an unpolarized beam is both necessary and sufficient. 
This asymmetry is comparable in size to the PVDIS asymmetry that has been measured at JLab to (2-3)\% 
precision~\cite{Wang:2014bba,Wang:2014guo}. 
We note that unlike the PVDIS asymmetry where the contribution from the $2g_{VA}^{eu}-g_{VA}^{ed}$ is quite small, 
the asymmetry in Eq.~(\ref{eq:asym}) arises fully from the couplings we wish to measure.

In practice, such a measurement will encounter both experimental and theoretical challenges. 
The DIS cross section difference between $e^-$ and $e^+$ scattering is subject to higher-order 
QED corrections. 
Box graphs describing two photon and photon-$Z$ boson exchange have to be included, combined with real photon
radiation to render the result infrared finite. 
The expected size of these corrections is $\mathcal{O}(\alpha/\pi)$, {\textit {i.e.}}\ at the $10^{-3}$ level, 
and without a logarithmic enhancement. 
The separation of the weak couplings $g_{AA}^{eq}$ requires theory predictions of such higher-order
effects at the level of 1\% or better. Theory techniques are ready to achieve this goal at the parton level including two-loop Feynman diagrams. A parton-level calculation is expected to be valid at large $Q^2$, but additional investigations will be required to improve our understanding of related uncertainties in the transition region towards the $Q^2$ values of JLab. It is beyond the scope of this article to perform a detailed study of these corrections. 
We trust that with dedicated efforts, future theory work will allow to control the associated uncertainties at the required level.  

\begin{figure}[t]
\includegraphics[width=0.47\textwidth]{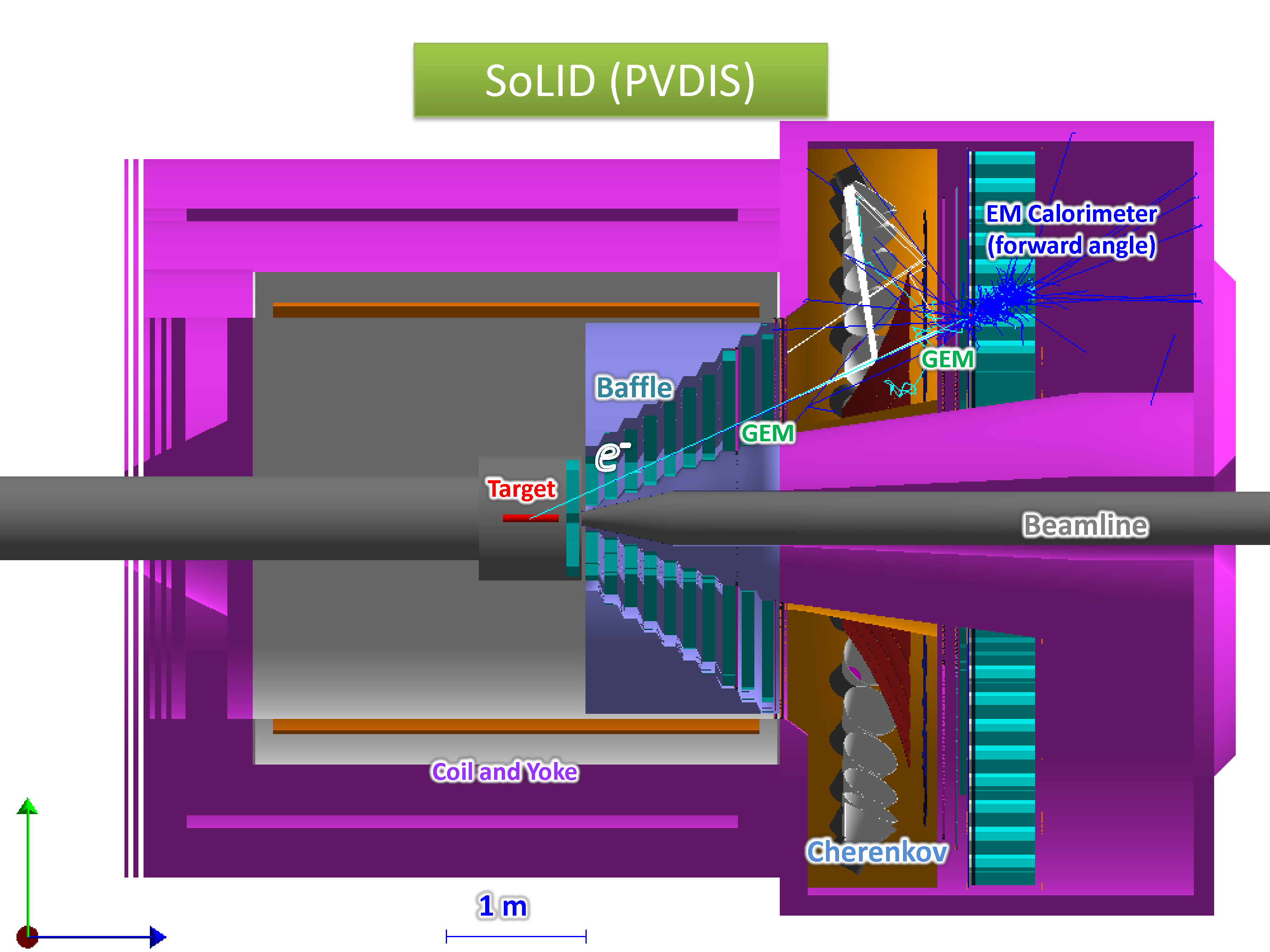}
\caption{SoLID in its PVDIS configuration. 
The electron (or positron) beam enters from the left and incident on a 40-cm long liquid target. 
Scattered electrons are detected by GEM chambers, a gas Cherenkov detector, and an electromagnetic 
calorimeter.  
A set of "baffles" (slitted shielding) can be used at high luminosity to reduce backgrounds, 
but may not be needed for the positron running discussed here.}
\label{fig:solid-pvdis}
\end{figure}

\section*{Feasibility of $A^{e^+e^-}_d$ Measurement at JLab}
We now consider to expose the planned Solenoid Large Intensity Device (SoLID)~\cite{Chen:2014psa} 
to a possible 11~GeV positron beam from the Continuous Electron Beam Accelerator Facility (CEBAF) at JLab in order to measure $A^{e^+e^-}_d$ on a deuteron target.
SoLID is a general-purpose, large-acceptance spectrometer that can handle the high luminosities of CEBAF. 
It is currently being planned for the experimental Hall~A for measurements of PVDIS, 
semi-inclusive DIS (SIDIS), and other physics processes including $J/\psi$ production and 
Deeply Virutual Compton Scattering (DVCS).  
The most suitable setup for our $A^{e^+e^-}_d$ measurement will be SoLID's PVDIS configuration 
shown in Fig.~\ref{fig:solid-pvdis}. 
To use the PVDIS setup for positron scattering, the polarity of the solenoid magnet will be reversed such that 
the acceptance of the scattered positrons can be kept as close to that of electrons as possible. 
Because of the large acceptance of SoLID, it is possible to measure $A^{e^+e^-}_d$ over a wide kinematic range 
and use the kinematic dependence of the asymmetry to isolate the electroweak contribution due to the $g_{AA}^{eq}$
from other competing effects.

\begin{figure}[t]
\includegraphics[width=0.48\textwidth]{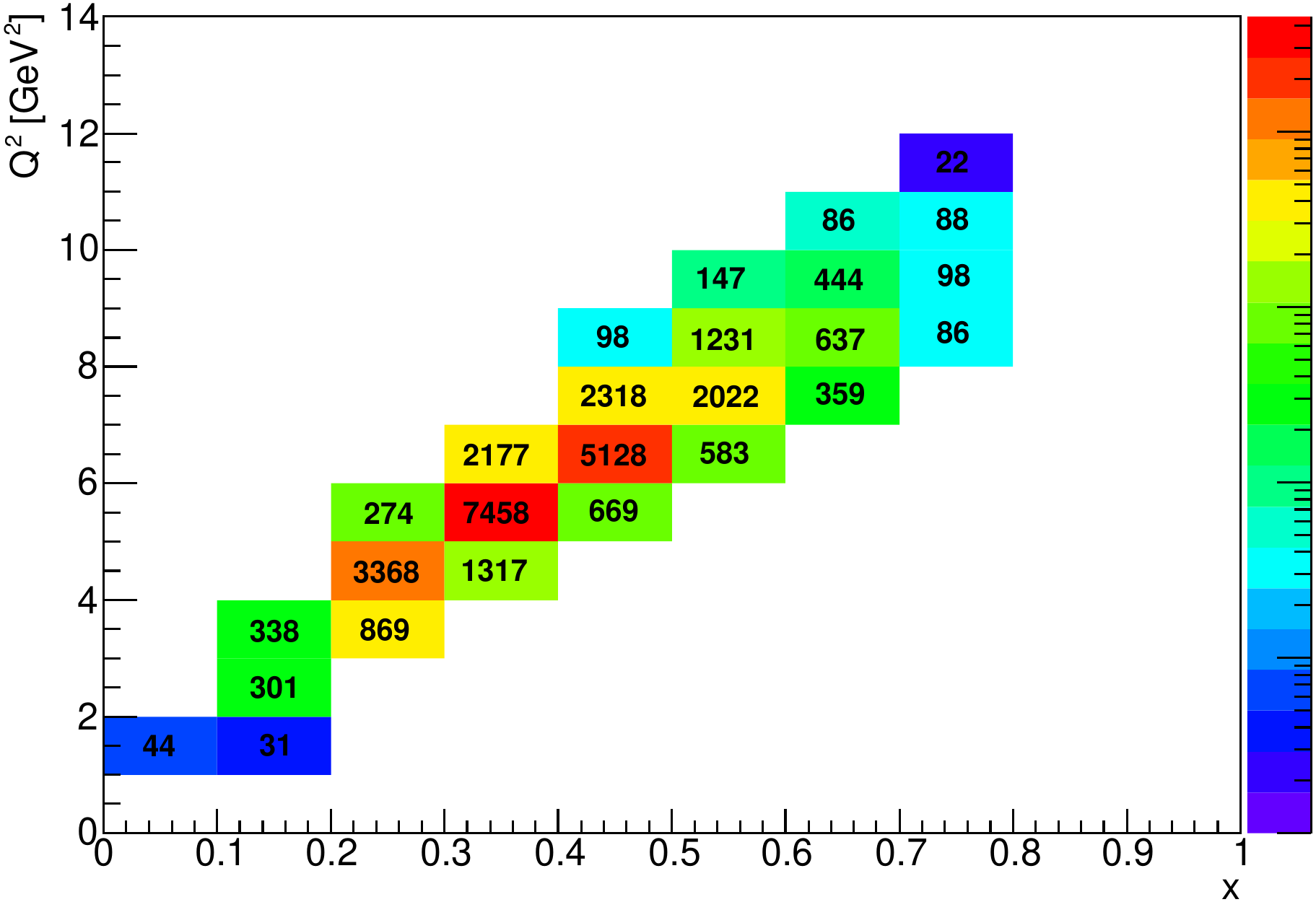}
\caption{Generated rates (in Hz) assuming the SoLID PVDIS configuration with a 40~cm long liquid deuterium 
target exposed to a 3~$\mu$A beam current, shown on a $(x,Q^2)$ grid ($Q^2$ in GeV$^2$). 
The cut on the invariant mass of the hadronic final state system of $W>2$~GeV and the $Q^2>1$~GeV$^2$ cut 
have been applied to ensure DIS. 
Baffles are not assumed.}
\label{fig:rate_1uA_nobaffle}
\end{figure}

\begin{figure}[t]
\includegraphics[width=0.48\textwidth]{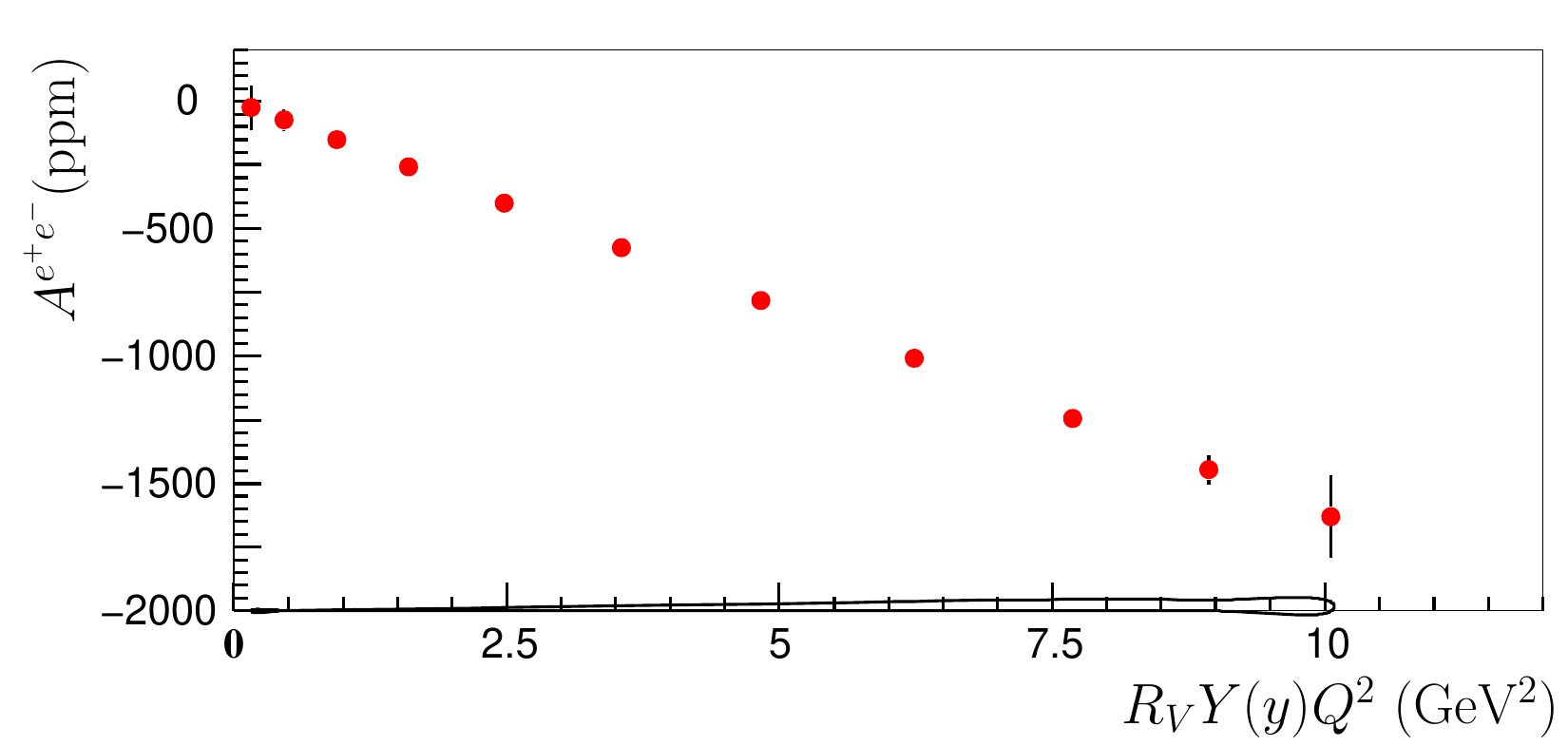}
\caption{Expected value and statistical uncertainty (in ppm) for a measurement of $A^{e^+e^-}_d$
using 20 days of 3~$\mu$A unpolarized beam at 100\% efficiency directed at a 40~cm long liquid deuterium 
target, interchanging between positrons and electrons.
SoLID is assumed in its PVDIS configuration without baffles. 
The horizontal band at $-2000$~ppm shows the total PDF uncertainty (see main text for details), 
multiplied by a factor of~10 for better visibility.}
\label{fig:c3fit}
\end{figure}

We studied the feasibility of a measurement of the asymmetry $A^{e^+e^-}_d$ using the simulated DIS rates from 
the SoLID pre-Conceptual Design Report~\cite{solid-pcdr} for a 3~$\mu$A positron beam incident on a 40~cm 
liquid deuterium target, as detailed in Fig.~\ref{fig:rate_1uA_nobaffle}. The value 3~$\mu$A for the beam current was chosen based on the present expectation of the positron source, demonstrated by the PEPPo (Polarized Electrons for Polarized Positrons) experiment~\cite{Cardman:2018svy}. 
The statistical uncertainty estimate assumes 20 days of beam time at 100\% efficiency as shown 
in Fig.~\ref{fig:c3fit}. 
We assumed that SoLID will be used in its PVDIS configuration with the baffles removed. 
If baffles were used, the required beam time would be about three times as long. 
The size of the asymmetry at each $Q^2$ point was calculated from Eq.~(\ref{eq:asym}), and both 
MMHT2014~\cite{Harland-Lang:2014zoa} (NLO120 grid) and CT18~\cite{Hou:2019efy} (CT18NLO grid) PDFs
averaged over the corresponding $x$ range. 
To estimate the total PDF uncertainty in the asymmetry calculation, we took the average of the $A^{e^+e^-}_d$ 
results from the two PDF sets as central values, and the half-differences between the two was added in 
quadrature to the uncertainty calculated using the PDF eigenvector sets.
This total PDF uncertainty varies from 10~ppm to below 1~ppm. 

The binned asymmetries can be fitted using 
\begin{eqnarray}
A^{e^+e^-}_d(Q^2) = a + b R_V \left[\frac{1-(1-y)^2}{1+(1-y)^2}\right] Q^2~, \label{eq:c3fit}
\end{eqnarray}
where $a$ and $b$ are free parameters. 
We found the variation $\Delta b=\pm 2.75$~ppm, resulting in a combined statistical plus PDF uncertainty 
of $\Delta (2g_{AA}^{eu}-g_{AA}^{ed}) = \pm 0.026$.
For an actual measurement, the fitting function~(\ref{eq:c3fit}) can be modified to account for other 
contributions, such as QED corrections, and 
differences in beam intensities and energies, and other experimental effects between 
$e^+$ and $e^-$ runs. 
Depending on the size of these experimental effects, the systematic uncertainty may dominate.

These estimates demonstrate that a measurement of the $g_{AA}^{eq}$ is possible using SoLID, 
provided one can switch between a positron and an electron beam. 
The differences between the $e^+$ and $e^-$ beams, from intensity, energy, position, direction, and spot 
size, need to be kept as small as possible and the effects studied carefully. 
One must also require that the difference in spectrometer response to electrons and positrons is below 
the statistical error, or can be corrected for. 
Note that while these requirements may look daunting to reach in practice, they had been achieved in 
the 96-day long CERN experiment~\cite{Argento:1982tq}, where a switch between $\mu^+$ and $\mu^-$ took place 
twice in each 12-day run period.
Care was taken to ensure that the $\mu^\pm$ data were collected at the same intensity such that many 
systematic effects cancel, and the spectrometer magnet was operating at full saturation so that the field 
could be reproduced to high precision with each polarity reversal, while a 40 meter long carbon target 
assured the required rates.

\section*{New Physics Mass Limit}
The energy scale, $\Lambda$, up to which new physics Beyond the SM (BSM) is testable, can be quantified
in terms of perturbations of the SM Lagrangian~(\ref{eq:L}), that is by replacements of the form,
\begin{eqnarray}
\frac{G_F}{\sqrt{2}} g_{ij} \rightarrow \frac{G_F}{\sqrt{2}} g_{ij} + \eta_{ij}^{q}\frac{4\pi}{(\Lambda_{ij}^{q})^2}\ ,
\label{eq:ciqmodified}
\end{eqnarray}
where $ij=AV,VA,AA$ and we assumed that the new physics is strongly coupled with a coupling $g$ given by $g^2 = 4\pi$. 
Without the specification of a particular BSM scenario, any combination of the $g_{AA}^{eq}$ can be modified, 
both with constructive or destructive interference with the SM, {\textit i.e.} $\eta_{ij}^{q} = \pm 1$. 
Therefore, a determination of the combination $2g_{AA}^{eu}-g_{AA}^{ed}$ through $A^{e^+e^-}_d$ puts constraints on 
$\Lambda_{AA}^u$ and $\Lambda_{AA}^d$. 
Of course, the sensitivity of a measurement of $2g_{AA}^{eu}-g_{AA}^{ed}$ is reduced (enhanced) in models 
in which the contributions to $g_{AA}^{eu}$ and $g_{AA}^{ed}$ are (anti-)correlated. 
In general, any BSM model will produce a shift in the combination
$\cos\alpha\, g_{AA}^{eu} + \sin\alpha\, g_{AA}^{ed}$ for some value of $\alpha$, to which a measurement of 
$2g_{AA}^{eu}-g_{AA}^{ed}$ has maximal (minimal) sensitivity for $\tan\alpha = -1/2$ $(+2)$.
The associated new physics scales which can be probed given a total uncertainty
$\Delta(\cos\alpha\, g_{AA}^{eu} + \sin\alpha\, g_{AA}^{ed})$, are given by,
\begin{eqnarray}
\Lambda(\alpha) = v \sqrt{\frac{8\pi}{\Delta (\cos\alpha\, g_{AA}^{eu} + \sin\alpha\, g_{AA}^{ed})}}\ ,
\end{eqnarray}
where $v=(\sqrt{2}G_F)^{-1/2}=246.22$~GeV is the Higgs vacuum expectation value setting the electroweak 
scale. 
Thus, in models predicting $\Delta g_{AA}^{eu} = 0$ ($\alpha=\pi/2$) or $\Delta g_{AA}^{ed} = 0$ ($\alpha=0$)
a constraint $\Delta(2g_{AA}^{eu}-g_{AA}^{ed}) = \pm 0.03$, can probe $\Lambda_{AA}^{d}$ up to 7.1~TeV or $\Lambda_{AA}^{u}$ up
to 10.1~TeV, respectively. 
The maximal $1\sigma$-sensitivity is given by $\alpha=\arctan(-1/2)$ and 
\begin{eqnarray}
\Lambda = v \sqrt{\frac{8\sqrt{5}\pi}{\Delta (2 g_{AA}^{eu} - g_{AA}^{ed})}} = 10.7~\rm TeV,
\end{eqnarray}
while the minimum $\Lambda=0$ is given by $\alpha=\arctan(2)$.

Any model predicting a significant effect in the $g_{AA}^{eq}$, while leaving the $g_{AV,VA}^{eq}$ unaltered, is presumably contrived or tuned; however, the $g_{AA}^{eq}$ are couplings independent of 
the $g_{AV,VA}^{eq}$, and their constraints on BSM physics are complementary. 
Conversely, if new physics is seen in the $g_{AV,VA}^{eq}$, it would be of paramount importance to measure 
the $g_{AA}^{eq}$, as well.

\section*{Summary}
We reviewed the current knowledge on the electron-quark effective couplings $g_{AV}^{eq}$, $g_{VA}^{eq}$ and $g_{AA}^{eq}$, 
all of which are accessible in charged lepton scattering. 
The $g_{AA}^{eq}$ can be extracted from $A^{e^+e^-}_d$, the DIS cross section asymmetry between a positron and an 
electron beam scattering off a deuterium target, assuming that other contributions can be reliably separated 
and that all experimental systematic uncertainties are under control.  
An exploratory calculation was performed to study the possibility of measuring $A^{e^+e^-}_d$ by exposing
the planned SoLID spectrometer at JLab to a future unpolarized positron beam. 
Our results are encouraging, mapping out the path to the first measurement of the electron-quark $g_{AA}^{eq}$ 
within a reasonable amount of beam time.  
A targeted precision of $\Delta(2g_{AA}^{eu}-g_{AA}^{ed}) = \pm 0.03$ would provide an order of magnitude stronger
constraint compared to the one on the muonic $g_{AA}^{\mu q}$ from CERN.
Detailed studies are underway on the control of experimental systematic uncertainties and QED corrections.

\begin{acknowledgements}
X.Z. was supported by the U.S. Department of Energy (DOE) Early Career Award SC00--03885 during the early 
stage of this work, and currently by the U.S.\ DOE under Award number DE--SC0014434. 
The work of J.E. and H.S. is supported by the German-Mexican research collaboration grant SP 778/4--1 (DFG) 
and 278017 (CONACyT).
\end{acknowledgements}

\appendix

\section{Neutral-current DIS asymmetries}
In this appendix, we derive asymmetries of lepton deep inelastic scattering off a nuclear target arising from 
the interference between electromagnetic and weak neutral current (NC) interactions. 
We are interested in the case of relatively small momentum transfer.

\subsection*{Neutral-current weak interaction Lagrangian}
\begin{figure}[t]
\includegraphics[width=0.48\textwidth]{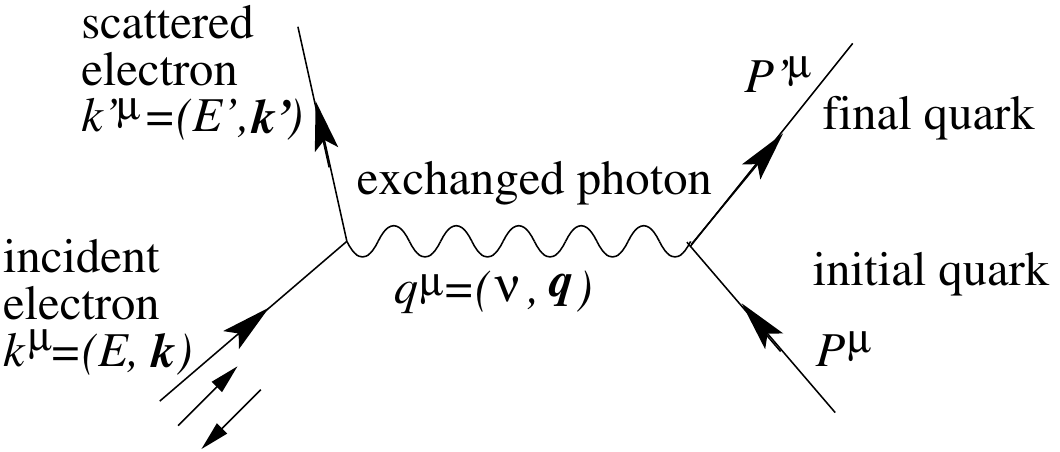}
\caption{One-photon exchange in electron-quark DIS. 
For the weak neutral current interaction, the photon is replaced by a $Z$ boson.}
\label{fig:eq}
\end{figure}

The kinematics of lepton-quark scattering is illustrated in Fig.~\ref{fig:eq}, where the photon-fermion vertex
is given by $-ie\gamma^\mu Q_f$ with $Q_f$ the fermion electric charge in units of $e=\sqrt{4\pi\alpha}$. 
Likewise, the $Z$-fermion vertex reads,
\begin{equation}
-i\frac{g}{2\cos\theta_W}\gamma^\mu \left(g_V^f - g_A^f\gamma^5\right)\ , 
\end{equation}
The matrix element for NC $eq$ scattering can now be written,
\begin{align}
\label{eq:MNCeq}
&\frac{1}{i}\mathcal{M}_{NC} = \bar l_f 
\left[ -i\frac{g}{2\cos\theta_W} \gamma^\mu \left(g_V^l - g_A^l\gamma^5\right) \right]l_i \\
&\left( -i\frac{g_{\mu\nu} - \frac{q_\mu q_\nu}{M_Z^2}}{q^2 - M_Z^2} \right) \bar q_f
\left[ -i\frac{g}{2\cos\theta_W} \gamma^\nu \left(g_V^q - g_A^q\gamma^5\right) \right]q_i\ , \nonumber
\end{align}
where $M_Z$ is the $Z$ boson mass, and $l_{i,f}$ and $q_{i,f}$ are the Dirac spinors for the initial and 
final state electrons and quarks, respectively. 
At the SM tree level, the gauge coupling $g$ is related to $G_F$,
\begin{equation}
\frac{G_F}{\sqrt{2}} = \frac{g^2}{8M_Z^2\cos^2\theta_W}\ ,
\end{equation}
so that for small momentum transfer, $\vert q^2\vert\ll M_Z^2$, and dropping the vector-vector interactions, 
the amplitude in Eq.~(\ref{eq:MNCeq}) derives from the neutral-current weak interaction Lagrangian
\begin{eqnarray}
&& \mathcal{L}_{int}^{NC} = \frac{G_F}{\sqrt{2}}
\left[ g_{VV}^{eq}\ \bar l_f \gamma^\mu l_i \bar q_f \gamma_\mu q_i\ \right. \nonumber \\
&+& \left. g_{VA}^{eq}\ \bar l_f \gamma^\mu l_i \bar q_f \gamma_\mu \gamma^5 q_i\ + 
g_{AV}^{eq}\ \bar l_f \gamma^\mu \gamma^5 l_i\ \bar q_f \gamma_\mu q_i \right. \nonumber \\
&+& \left. g_{AA}^{eq}\ \bar l_f \gamma^\mu \gamma^5 l_i\ \bar q_f \gamma_\mu \gamma^5 q_i \right], 
\label{eq:MNC}
\end{eqnarray}
where we used the same symbols for Dirac spinor fields and coefficient functions.

\subsection*{One-photon exchange amplitude and $Z$--$\gamma^*$ interference}
For incoming leptons with helicity $h$ (the case of anti-leptons will be treated separately), we have 
the one-photon exchange amplitude,
\begin{equation}
\mathcal{M}_{\gamma}^h = - Q_l Q_q \frac{4\pi\alpha}{q^2}
(\bar l_f\gamma^\mu P_h l_i)(\bar q_f\gamma_\mu q_i)\ ,\label{eq:Mgamma1}
\end{equation}
where,
\begin{equation} 
P_h \equiv \frac{1+h\gamma^5}{2}\ ,
\end{equation}
is the projection operator for right-handed ($h = +1$) and left-handed ($h = -1$) leptons. 
The cross section of the electromagnetic process is,
\begin{eqnarray}
\sum_{\mathrm{spins}}\vert\mathcal{M}_h^\gamma\vert^2 &=& 
16 Q_l^2 Q_q^2\left(\frac{4\pi\alpha}{q^2}\right)^2 \nonumber \\
&\times& \left[ (Pk)(P'k') + (Pk')(P'k) \right].
\end{eqnarray}
For the weak neutral current interaction we have,
\begin{eqnarray}
\mathcal{M}_{NC}^h &=& \sqrt{2}G_F \left[\bar l_f \gamma^\mu\left(g_V^l - g_A^l\gamma^5\right)P_h l_i\right] 
\nonumber \\
&\times& \left[ \bar q_f\gamma_\mu\left(g_V^q-g_A^q\gamma^5\right)q_i \right], \label{eq:MZ4}
\end{eqnarray}
so that the interference term is given by,
\begin{eqnarray}
&&\left(\mathcal{M}_\gamma^h\right)^\ast \mathcal{M}_{NC}^h = - \frac{4\sqrt{2}\pi G_F\alpha}{q^2}Q_l Q_q
(\bar l_i P_{-h} \gamma^\mu l_f) \times \\
&&\left[\bar l_f\gamma^\nu(g_V^l - g_A^l \gamma^5) P_h l_i \right]
(\bar q_i\gamma_\mu q_f)\left[\bar q_f\gamma_\nu(g_V^q - g_A^q \gamma^5) q_i \right]. \nonumber
\end{eqnarray}
Averaging over initial and summing over the final spin states,
\begin{eqnarray}
\sum_{\mathrm{spins}} && \left(\mathcal{M}_{\gamma}^{h}\right)^\ast \mathcal{M}_{NC}^h = 
- \frac{64\sqrt{2}\pi G_F\alpha}{q^2}Q_l Q_q \times \label{eq:MM-lq-kP} \\ \nonumber
&& \left\{ (kP)(k'P') \left[ g_V^l g_V^q - h g_A^l g_V^q - h g_V^l g_A^q + g_A^l g_A^q \right] + \right. \\
&& \left. (kP')(k'P) \left[ g_V^l g_V^q - h g_A^l g_V^q + h g_V^l g_A^q - g_A^l g_A^q \right]  \right\} \, . 
\nonumber
\end{eqnarray}

Likewise, denoting the anti-lepton coefficient functions by $v_{i,f}$, for incoming anti-leptons with helicity
$h$ the leptonic currents in Eq.~(\ref{eq:Mgamma1}) and~(\ref{eq:MZ4}) are now $(\bar v_iP_h\gamma^\mu v_f)$ 
and $\left[\bar v_iP_h \gamma^\mu\left(g_V^l-g_A^l\gamma^5\right)v_f\right]$, respectively.
Eq.~(\ref{eq:MM-lq-kP}) then also applies for anti-leptons as long as one substitutes $h\to -h$ and 
$g_A^q\to -g_A^q$.
  
For anti-quarks at the $\gamma$ or $Z^0$ vertex one substitutes analogously the quark-coefficient functions
by those of anti-quarks, with the result that Eq.~(\ref{eq:MM-lq-kP}) applies when one replaces 
$g_A^q\to -g_A^q$ in the case of lepton scattering. 
Finally, for anti-lepton scattering off anti-quarks one needs to substitute $h\to -h$ in 
Eq.~(\ref{eq:MM-lq-kP}).

\subsection*{Electroweak neutral current cross section asymmetries}
In DIS one has to combine the scattering cross sections for all quarks in the target with weights according
to the PDFs.
From the definition in Eq.~(\ref{eq:Asym}) we have,
\begin{eqnarray}
A_{RL}^{e^\pm} &=& \frac{\vert\mathcal{M}_Z+\mathcal{M}_\gamma\vert_{h=+\vert\lambda\vert}^2 - 
\vert\mathcal{M}_Z + \mathcal{M}_\gamma\vert_{h=-\vert\lambda\vert}^2}
{\vert\mathcal{M}_Z + \mathcal{M}_\gamma\vert_{h=+\vert\lambda\vert}^2 + 
\vert\mathcal{M}_Z + \mathcal{M}_\gamma\vert_{h=-\vert\lambda\vert}^2} \nonumber \\
&\approx& \frac{(\mathcal{M}_\gamma^\ast  \mathcal{M}_Z)_{h=+\vert\lambda\vert} - 
(\mathcal{M}_\gamma^\ast  \mathcal{M}_Z)_{h=-\vert\lambda\vert}}{\vert\mathcal{M}_\gamma\vert^2}\ .
\end{eqnarray}  
If we now use the previous results, inserting the Mandelstam variables neglecting lepton and quark masses, 
{\textit {i.e.}}, $s \equiv (k+P)^2 = 2 kP = 2 k'P'$ with $P_{\rm quark} = x P_{\rm nucleon}$,
and likewise $u \equiv (k-P')^2 = -2 kP' = -2 k'P= -(1-y) s$,
\begin{eqnarray}
A^{e^-}_{RL} &=& \vert\lambda\vert \frac{G_F Q^2}{2\sqrt{2}\pi\alpha}
\left\{ \frac{\sum q(x,Q^2)Q_q g_{AV}^{eq}[1+(1-y)^2]}{\sum q(x,Q^2) Q_q^2[1+(1-y)^2]} \right. \nonumber \\
&+& \left. \frac{\sum q(x,Q^2)Q_q g_{VA}^{eq}[1-(1-y)^2]}{\sum q(x,Q^2) Q_q^2[1+(1-y)^2]} \right\}\ ,
\end{eqnarray}
where $\lambda$ is the beam polarization and $q(x,Q^2)$ are PDFs. 
The value $Q_l=-1$ for electron scattering, $Q^2 \equiv -q^2$, and the definitions of $g_{AV}^{eq}$ and $g_{VA}^{eq}$ 
were also used. 
We note that for the antiquark contributions the couplings $g_A^q$ (and therefore all $g_{VA}^{eq}$) appear with 
an extra minus sign.
For a proton target we obtain,
\begin{eqnarray}
A^{e^-}_{RL,p} &=&  \vert\lambda\vert \frac{3 G_F Q^2}{2\sqrt{2}\pi\alpha(4U^++D^+)}
\left[(2U^+ g_{AV}^{eu} - D^+ g_{AV}^{ed}) \right. \nonumber \\
&+& \left. Y(2 u_V g_{VA}^{eu}-d_V g_{VA}^{ed}) \right]\ ,
\end{eqnarray}
where we have further abbreviated 
$U^+ \equiv u^+ + c^+$ and $D^+ \equiv d^+ + s^+$, and have assumed $c=\bar c$ and
$s=\bar s$ (and thus $c_V=s_V=0$). The function $Y$ is defined in Eq.~(\ref{eq:Y}) and we will omit its dependence on $y$ hereafter. 
For deuterium or other isoscalar targets (ignoring nuclear effects), we substitute $u\to u+d$ and $d\to u+d$ 
in the expression for $A^{e^-}_{RL,p}$ above, and assume that $c$ and $s$ are the same in the proton and 
the neutron: 
\begin{eqnarray}
A^{e^-}_{RL,d} &=& \vert\lambda\vert \frac{3 G_F Q^2}{2\sqrt{2}\pi\alpha(5 + 4 R_C + R_S)}
\left\{ \left[ 2 (1 + R_C) g_{AV}^{eu} \right. \right. \nonumber \\
&-&\left. \left. \hspace*{-2mm}(1 + R_S) g_{AV}^{ed} \right] +Y (2 g_{VA}^{eu} - g_{VA}^{ed}) R_V \right\}\ .
\end{eqnarray}
Ignoring the heavier $s$ and $c$ quarks, these expressions simplify further,
\begin{eqnarray}
A^{e^-}_{RL,p} &\approx& \vert\lambda\vert \frac{3 G_F Q^2}{2 \sqrt{2}\pi\alpha (4u^+ + d^+)}
\left[ (2 u^+ g_{AV}^{eu} - d^+ g_{AV}^{ed}) \right. \nonumber \\
&+& \left. Y (2 u_V g_{VA}^{eu} - d_V g_{VA}^{ed}) \right], \\
A^{e^-}_{RL,d} &\approx& \vert\lambda\vert \frac{3 G_F Q^2}{10 \sqrt{2}\pi\alpha}
\left[ (2 g_{AV}^{eu} - g_{AV}^{ed}) \right. \nonumber \\ 
&+& \left. R_V Y (2g_{VA}^{eu} - g_{VA}^{ed}) \right].
\end{eqnarray}

We now turn to the calculation of $A^{e^+e^-}_{RL}$,
\begin{eqnarray}
A^{e^+e^-}_{RL} &=& \frac{\vert\mathcal{M}_Z^{e^+}+\mathcal{M}_\gamma^{e^+}\vert^2_{h=+\vert\lambda\vert}
-\vert\mathcal{M}_Z^{e^-}+\mathcal{M}_\gamma^{e^-}\vert^2_{h=-\vert\lambda\vert}}
{\vert\mathcal{M}_Z^{e^+}+\mathcal{M}_\gamma^{e^+}\vert^2_{h=+\vert\lambda\vert}
+\vert\mathcal{M}_Z^{e^-}+\mathcal{M}_\gamma^{e^-}\vert^2_{h=-\vert\lambda\vert}} \nonumber \\ 
&\approx&\frac{(\mathcal{M}_\gamma^{e^+}\mathcal{M}_Z^{e^+})_{+\vert\lambda\vert}
- (\mathcal{M}_\gamma^{e^-}\mathcal{M}_Z^{e^-})_{-\vert\lambda\vert}}{\vert\mathcal{M}_\gamma\vert^2}\ ,
\end{eqnarray}
where the approximation is valid for $Q^2 \ll M_Z^2$. 
For a nuclear target,
\begin{equation}
A^{e^+e^-}_{RL} = 
\frac{G_F Q^2Y\sum_q q_V Q_q (\vert\lambda\vert g_{VA}^{eq} - g_{AA}^{eq})}{2\sqrt{2}\pi\alpha \sum_q q^+ Q_q^2}\ . 
\label{eq:ARL}
\end{equation}
and likewise,
\begin{eqnarray}
A^{e^+e^-}_{RR} = 
\frac{G_F Q^2\sum_q Q_q [-\vert\lambda\vert q^+ g_{AV}^{eq}-q_V Y g_{AA}^{eq}]}{2\sqrt{2}\pi\alpha\sum_q q^+ Q_q^2}\ .
\label{eq:ARR}
\end{eqnarray}
By substituting $\vert\lambda\vert\to -\vert\lambda\vert$ one can obtain $A^{e^+e^-}_{LR}$ from 
Eq.~(\ref{eq:ARL}), and $A^{e^+e^-}_{LL}$ from Eq.~(\ref{eq:ARR}). 
For $A^{e^+e^-}$ one can use (\ref{eq:ARL}) or (\ref{eq:ARR}) and set $\vert\lambda\vert=0$.
For the proton,
\begin{eqnarray}
A^{e^+e^-}_{RL,p} &=& \frac{3 G_F Q^2 Y}{2\sqrt{2}\pi\alpha(4 U^+ + D^+)} 
\left[\vert\lambda\vert (2 u_V g_{VA}^{eu} - d_V g_{VA}^{ed}) \right. \nonumber \\
&-& \left. ({2} u_V g_{AA}^{eu} - d_V g_{AA}^{ed}) \right]\ ,
\end{eqnarray}
\begin{eqnarray}
A^{e^+e^-}_{RR,p} &=& \frac{3 G_F Q^2}{2\sqrt{2}\pi\alpha({4}U^+ + D^+)}
\left[-\vert\lambda\vert (2 U^+ g_{AV}^{eu} - D^+ g_{AV}^{ed}) \right. \nonumber \\
&-& \left. Y ({2}u_Vg_{AA}^{eu}-d_Vg_{AA}^{ed}) \right]\ ,
\end{eqnarray}
and for the deuteron,
\begin{eqnarray}
\label{eq:ARL-d}
A^{e^+e^-}_{RL,d} &=& \frac{3 G_F Q^2 Y R_V}{2\sqrt{2}\pi\alpha(5 + 4 R_C + R_S)} 
\left[\vert\lambda\vert (2 g_{VA}^{eu} - g_{VA}^{ed}) \right. \nonumber \\
&-& \left. (2 g_{AA}^{eu} - g_{AA}^{ed}) \right]\ ,
\end{eqnarray}
\begin{eqnarray}
A^{e^+e^-}_{RR,d} &=& \frac{3 G_F Q^2}{2\sqrt{2}\pi\alpha(5 + 4 R_C + R_S)}
\left\{ -\vert\lambda\vert \left[ 2 (1 + R_C) g_{AV}^{eu} \right. \right. \nonumber \\
&-& \left. \left. (1 + R_S) g_{AV}^{ed} \right] - Y R_V (2 g_{AA}^{eu} - g_{AA}^{ed}) \right\}\ .
\end{eqnarray}
Finally, if only $u$ and $d$ quarks are included,
\begin{eqnarray}
A^{e^+e^-}_{RL,p} &\approx& \frac{3 G_F Q^2 Y}{2\sqrt{2}\pi\alpha(4 u^+ + d^+)} 
\left[ \vert\lambda\vert (2 u_V g_{VA}^{eu} - d_V g_{VA}^{ed}) \right. \nonumber \\
&-& \left. (2 u_V g_{AA}^{eu} - d_V g_{AA}^{ed}) \right]\ ,
\end{eqnarray}
\begin{eqnarray}
A^{e^+e^-}_{RR,p} &\approx& \frac{3 G_F Q^2}{2\sqrt{2}\pi\alpha(4 u^+ + d^+)} 
\left[ -\vert\lambda\vert (2 u^+ g_{AV}^{eu} - d^+ g_{AV}^{ed}) \right. \nonumber \\
&-& \left. (2 u_V g_{AA}^{eu} - d_V g_{AA}^{ed}) Y \right]\ ,
\end{eqnarray}
and
\begin{eqnarray}
\label{eq:ARL-d1}
A^{e^+e^-}_{RL,d} &\approx& \frac{3 G_F Q^2 Y R_V}{10\sqrt{2}\pi\alpha} 
\left[ \vert\lambda\vert (2 g_{VA}^{eu} - g_{VA}^{ed}) \right. \nonumber \\
&-& \left. (2 g_{AA}^{eu} - g_{AA}^{ed}) \right]\ . 
\end{eqnarray}
\begin{eqnarray}
A^{e^+e^-}_{RR,d} &\approx& \frac{3 G_F Q^2}{10\sqrt{2}\pi\alpha} 
\left[ -\vert\lambda\vert (2 g_{AV}^{eu} - g_{AV}^{ed}) \right. \nonumber \\
&-& \left. Y R_V (2 g_{AA}^{eu} - g_{AA}^{ed}) \right]\ .
\end{eqnarray}
The asymmetry measured at CERN on $^{12}$C was,
\begin{eqnarray}
A^{\mu^+\mu^-}_{LR,C} &=& - \frac{3 G_F Q^2 Y R_V}{2\sqrt{2}\pi\alpha(5 + 4 R_C + R_S)}
\left[ (2 g_{AA}^{\mu u} - g_{AA}^{\mu d}) \right. \nonumber \\
&+& \left. \vert\lambda\vert (2 g_{VA}^{\mu u} - g_{VA}^{\mu d}) \right]\ ,
\end{eqnarray}
while for SoLID we can use unpolarized beams to allow for higher intensities and measure on the deuteron,
\begin{eqnarray}
A^{e^+e^-}_d = - \frac{3 G_F Q^2 Y}{2\sqrt{2}\pi\alpha}  \frac{R_V (2 g_{AA}^{eu} - g_{AA}^{ed})}{5 + 4 R_C + R_S}\ .
\end{eqnarray}

\end{document}